\documentclass{aa}  

\usepackage{graphicx}
\usepackage{txfonts}
\usepackage{hyperref}
\renewcommand{\arraystretch}{1.4}

\usepackage{tikz}

\begin{document}

   \title{Separation of polarized dust emission in Planck observations with Scattering Transforms}


 \author{
    A. Tsouros\inst{1} \and
    E. Russier\inst{2, 3, 4}
    E. Allys\inst{1} \and
    C. Auclair \and
    F. Boulanger \inst{1} \and
    J. Delabrouille\inst{4, 2}
}

   \institute{Laboratoire de Physique de l’École Normale Supérieure, ENS, Université PSL, CNRS, Sorbonne Université, Université Paris Cité,
75005 Paris, France
            \and
             Lawrence Berkeley National Laboratory, 1 Cyclotron Road, Berkeley, CA 94720, USA
            \and
            Department of Astronomy and Astrophysics, University of Chicago, 5640 S. Ellis Ave.,
Chicago, IL 60637, USA
            \and
            CNRS-UCB International Research Laboratory, Centre Pierre Binétruy, IRL 2007, CPB-IN2P3, Berkeley, CA 94720, USA
             }



  \abstract
{Polarized dust emission is a major astrophysical foreground contaminant for the measurement of cosmic microwave background polarization, which must be accurately measured to look for the faint primordial polarization B-modes of inflationary origin. The best currently available maps, obtained from \textit{Planck} space mission data, are noise-dominated in the high Galactic latitude regions that are most relevant for CMB observations.}
{The goal of this work is to obtain better dust polarization maps from \textit{Planck} observations, by exploiting both the dependence between polarization and total intensity, as well as the non-Gaussian filamentary structure of the dust emission.}
{To this effect, we use scattering transforms, which provide a stable and interpretable representation of complex non-Gaussian textures, allowing for a data-driven analysis approach requiring no explicit priors on dust. The analysis is performed locally on Cartesian patches of sky, where Stokes linear polarization parameters, redefined in a local reference frame, are modeled as the sum of a signal of interest and a nuisance term. Using multiple realizations of the random nuisance term, we recover the polarized dust maps by minimizing a composite objective function that enforces multiple statistical constraints in scattering space.}
{The proposed algorithm reconstructs maps of polarized dust emission whose statistics are consistent with those expected from the \textit{Planck} data once random nuisance realizations are added. This is confirmed in a validation test using a high signal-to-noise sky region as a test case. Comparisons with existing dust polarization maps and models show that our approach better recovers small-scale polarized dust emission, and that our reconstructed power and cross-spectra closely match those of the dust polarization maps. A second set of maps that deterministically reproduce the features of the dust polarized emission is also produced.}
   {}


   \maketitle
%

\section{Introduction}

The Planck space mission, launched by ESA in 2009, has provided the astronomical community with an unprecedented view of the microwave sky emission in nine frequency bands ranging from 30~GHz to 857~GHz. Seven of the Planck mission frequency channels, between 30~GHz and 353~GHz, were polarization-sensitive and have mapped a mixture of polarized emission from the cosmic microwave background (CMB) and from the interstellar medium (ISM) of our own Galaxy, the Milky Way \citep{2020A&A...641A...1P}. The Planck space mission data products have been made available to the scientific community in the Planck Legacy Archive at ESA.~\footnote{https://pla.esac.esa.int/pla/\#home}

The next generation of ground-based CMB observations is now focusing on precisely measuring CMB polarization, which is expected to carry the tiny signature of primordial gravitational waves generated during a phase of cosmic inflation~\citep{2016ARA&A..54..227K}. However, detecting polarized CMB is difficult because it is subdominant at all frequencies compared to polarized Galactic emission. The main Galactic contaminant above $\sim$70~GHz is polarized dust emission from the Galactic ISM, which arises from the preferential alignment of elongated dust grains perpendicular to the Galactic magnetic field. Below $\sim$70~GHz, polarized synchrotron emission from relativistic electrons dominates.

Traditionally, these different signals have been separated by exploiting their different emission laws as a function of frequency \citep{2009LNP...665..159D}.
This requires observations in frequency bands that ideally cover most of the 30 to 300~GHz frequency range. However, observations above~220 GHz are challenging from ground-based observatories, because of strong fluctuating atmospheric emission and absorption. This motivates the use of external templates of polarized dust emission to help with the analysis of future ground-based observations. Such maps have been made available as part of the Planck mission data releases~\citep{2020A&A...641A...4P}, but the published maps are either strongly filtered or dominated by noise in regions of the sky with the lowest polarized dust emission that are of greatest interest for current and future CMB observations. This paper aims to demonstrate how higher-quality polarized dust maps can be constructed from Planck data.


To do so, a promising new direction is to rely on recently developed Scattering Transforms (ST)~\citep{bruna2013invariant}. STs provide a mathematically grounded set of low-variance summary statistics that efficiently capture the non-Gaussian features of complex physical fields. Inspired by convolutional neural networks but defined analytically, they characterize interactions between oriented spatial scales through cascades of wavelet convolutions and nonlinear operations, as well as covariance estimates, producing compact and interpretable descriptors that go beyond traditional power-spectrum analyses. Because they require no training and can be estimated from limited data, STs have proven highly effective across a range of astrophysical and cosmological applications (\citealt{Allys2019}; \citealt{Regaldo2020}; \citealt{Cheng2021a}; \citealt{Valogiannis2022}); \citealt{lei2023probing}; \citealt{hothi2024wavelet}.

In addition to classification and parameter inference, ST have also been used to construct generative models of physical fields, in a maximum entropy framework. These quantitatively realistic models, constrained from the ST statistics themselves, can even be constructed from a single realization of the process under study~\citep{Allys2020, Cheng2024,Mousset2024}. They enable the synthesis of new, statistically consistent realizations that preserve the multiscale, non-Gaussian texture of various data types without any prior knowledge of the underlying physics or the need for extensive training sets. In~\cite{niall2022}, it was shown in particular that a ST-based generative model constructed from a single polarized dust patch could be used to train a neural network capable of separating CMB B-modes from Galactic dust emission, underscoring the potential of this framework for Galactic emission modeling. 

The ability to sample new realizations, which is done using a pixel-based gradient descent under ST constraints~\citep{bruna2019multiscale}, has been extended to component separation tasks. 
In~\citet{Regaldo2021}, this approach was introduced to separate the Galactic dust emission from instrumental noise in Planck 353\,GHz polarization maps. It has then been extended in~\citet{Delouis2022} to full-sky data, dust polarization maps that remain statistically reliable even at scales where the dust power 
is significantly below the noise level. 
More recently, \citet{Auclair2024} applied such a ST-based component separation to Herschel observations, showing that two distinct non-Gaussian processes---Galactic dust emission and the cosmic infrared background--- could be 
statistically separated from observational data alone, within a single-frequency framework. 
These approaches were also complemented by machine learning when sufficient data was available. This allowed an unsupervised ST modeling of components from unlabeled mixtures to be performed using Variational Auto-Encoders (VAEs), for both seismic data~\citep{siahkoohi2023unearthing,siahkoohi2023martian} and radio observations of the different phases of the interstellar medium~\citep{lei2025}.
These results demonstrate the great promise of such approaches. 



In this paper, we extend previous ST-based component separation approaches to construct maps of polarized dust emission at 353\,GHz with significantly improved angular resolution and reduced noise contamination compared to current state-of-the-art methods. 
The key novelties of our approach include the use of the most recent state-of-the-art ST statistics, the incorporation of information from 857\,GHz intensity maps to inform and constrain the separation of $Q$ and $U$ polarized components, 
making use of local polarization reference axes,
and the extensive use of complementary constraints. 



The paper is structured as follows. Section~\ref{sec:data} presents the observational data used in this work, as well as their preliminary processing. Section~\ref{sec:formalism} presents the mathematical formulation of the problem and the optimization scheme used to recover the component-separated maps. Section~\ref{sec:results} presents the results of our component separation algorithm on a sky patch and evaluates its performance using an independent validation patch. Finally, Section~\ref{sec:conc} presents our conclusions. 


\vspace{-0.2cm}
\section{Data} \label{sec:data}
\subsection{Set of maps used and preprocessing}

This paper focuses on producing a de-noised dust polarization map in the 353~GHz Planck frequency band. 

Polarized dust emission maps have been previously published by the Planck collaboration \citep{2020A&A...641A...4P}. Maps obtained with the Commander \citep{commander}, GNILC \citep{Remazeilles_2011}, and SMICA \citep{2003MNRAS.346.1089D,2008ISTSP...2..735C} methods have been made available on the Planck Legacy Archive. However, these maps, obtained in analyses for which the CMB was the primary objective, are either at degraded angular resolution (GNILC) or still significantly contaminated by residual noise (SMICA and Commander).

In the following, we use the Planck NPIPE PR4 maps~\citep{npipe2020}, which have the highest signal-to-noise ratio for dust polarization among the available Planck maps. 
These maps contain a small, but not negligible contribution from CMB polarization that we reduce by subtracting an estimate of CMB polarization obtained by a multivariate Wiener filter of the Planck $I$, $E$ and $B$ maps.\footnote{We use the SMICA CMB maps from the Planck PR3 data release.~\citep{smica_2020} and the theoretical CMB $I$, $E$ and $B$ auto and cross spectra, to evaluate the Wiener-filtered CMB maps.}
We also convolve the maps to a resolution of a Gaussian beam with an effective full width at half maximum (FWHM) of 10~arcmin, assuming an FWHM of 4.76~arcmin for the original PR4 maps, and convert units to MJy/sr. 
The obtained $Q$ and $U$ polarization maps are called $d_Q$ and $d_U$.

To perform the component separation, we rely on a model of the other signals than dust at this frequency, the instrumental noise and the CMB residual, the sum of which we call the contamination. To do so, we construct an ensemble of one hundred 353~GHz $Q$ and $U$ contamination maps, as described in App.~\ref{app:input_data}. These maps, which are also convolved to a resolution of 10~arcmin, are called $\left\{c_{Q,i},c_{U,i}\right\}$, for $i$ between 1 and 100.

In this paper, we also rely on the total intensity emission of dust, which is a good indicator of the location and shape of the dust structures that contribute to 353~GHz dust polarization maps. The Planck-HFI 353~GHz to 857~GHz maps are good tracers of this emission. Among those, we select the Planck NPIPE PR4 857~GHz map, which has the best signal-to-noise ratio, and is the least contaminated by cosmic infrared background (CIB) and CMB intensity fluctuations relatively to dust emission (see Fig.~\ref{fig:data}). Minimal preprocessing is performed on this map to detect and subtract emission from external galaxies and from the dense regions in the interstellar medium in the Milky Way. We also re-adjust the zero-level of the map by fitting a cosecant law to map emission at Galactic latitudes $|b|>10^\circ$.
This last map, which is also at at 10~arcmin, is called $d_I$.

We compare to the state-of-the-art polarized foreground maps, GNILC PR3~\citep{smica_2020}, which correspond to the Generalized Needlet Internal Linear Combination Method applied to the Planck 3rd data release. GNILC is a multi-frequency, scale-dependent component separation technique that operates in needlet space and uses local covariance estimates to isolate the foreground subspace while suppressing noise and CMB contamination \citep{Remazeilles_2011}. By adapting to spatial and angular-scale variations of galactic emission, GNILC provides polarized foreground emission templates that form the basis of current state-of-the-art dust models, including those implemented in PySM~\citep{Borrill_2025}.

\subsection{Selection of square patches} \label{sec:patches}

\begin{figure}[h!]
    \centering
    \includegraphics[width=1\linewidth]{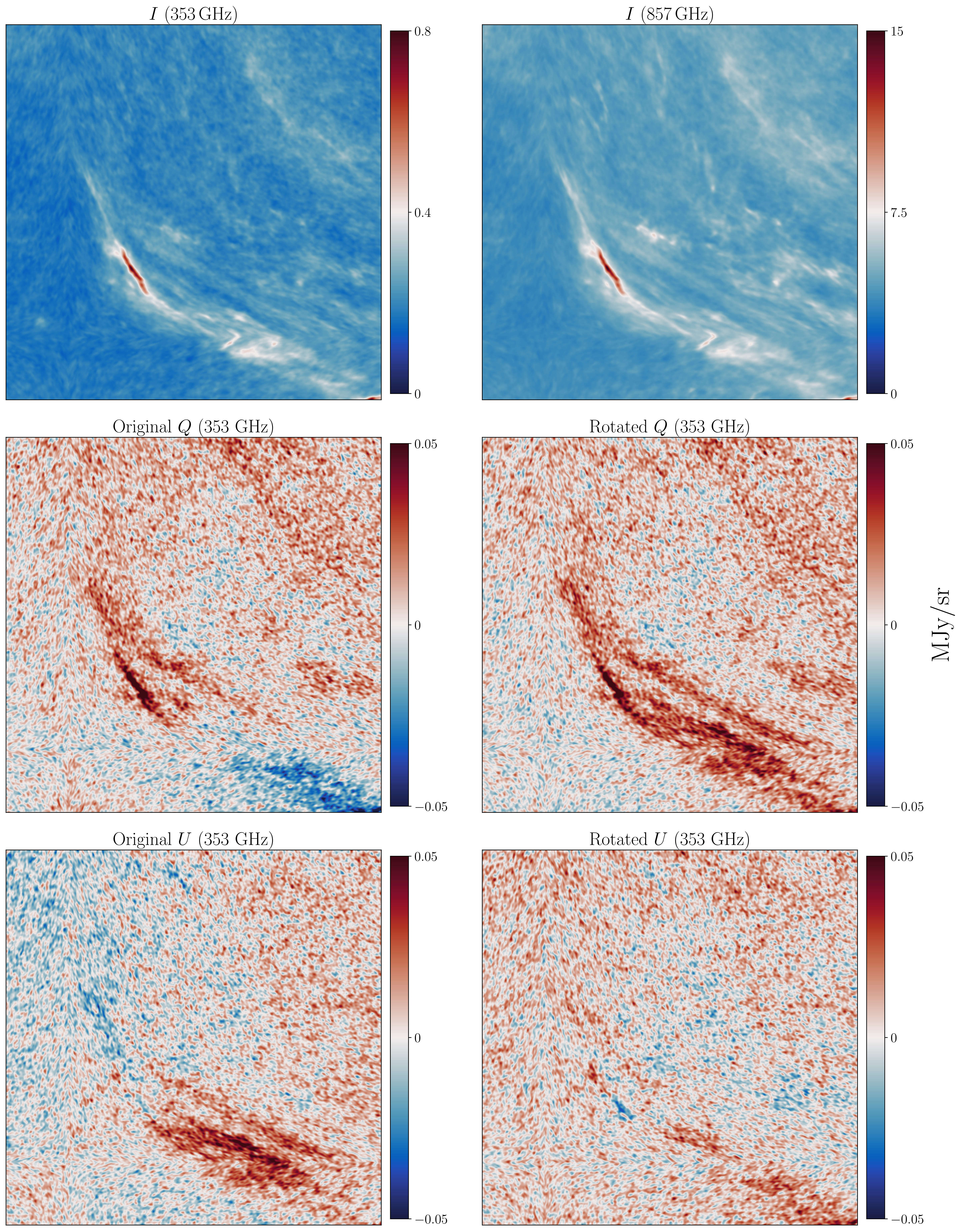}
    \caption{Signal maps of the patch of interest in our work centered at (l, b) = (315, 78.): Top left: I 353~GHz, Top right: I 857~GHz, Middle left: Q 353~GHz before polarization rotation, Middle right: Q 353~GHz after polarization rotation, Bottom left: U 353~GHz before polarization rotation, Bottom right: U 353~GHz after polarization rotation. In this paper, we only use the data from the right column.
    \vspace{-0.3cm}}
    \label{fig:data}
\end{figure}

In this paper, we only work on square patches, 
whose extraction from HEALpix maps is discussed in App.~\ref{app:HEALPix}. We also discuss in this appendix the geometric convention used for the definition of Stokes parameters consistently for each map, which solves the coordinate dependence problem of the original Stokes $Q$ and $U$ maps which creates artificial gradients along dust filaments as shown in Fig.~\ref{fig:data}. 
The size of the square patches is $384 \times 384$ pixels, and the linear scale of each pixel is $\sim 3.4$ arcmin.

We apply our component separation algorithm to two distinct regions. The first region is where we produce a dust map with an improved signal-to-noise ratio (SNR) and angular resolution. The second region, which already has a high SNR, is only used for validation. Specifically, we apply our component separation algorithm to this region after contaminating it to an SNR similar to that of the first region we study.

The first region corresponds to a patch centered at $(l, b) = (315^\circ, 78.3^\circ)$, chosen because it is at high Galactic latitude, and because it contains a significant dust filament located close to the center of the patch. Henceforth, we will refer to this patch as the \emph{North patch}. 
The $I$, $Q$, and $U$ of this region can be seen in Fig.~\ref{fig:data}. The left column shows the $I$ map at 353~GHz, and the original $Q$ and $U$ polarized map before the redefinition of the Stokes parameters. The right column corresponds to the fields that will be used below: $d_I$ for the $I$ field at 857 GHz, and $d_Q$, $d_U$, the rotated $Q$ and $U$ maps at 353~GHz.

Incidentally, one sees that this is a region where the original $Q$ and $U$ reference system rotates strongly across the patch. Indeed, by looking along the filament in the original $Q$ and $U$ maps, we see an anti-correlation along the filament in both cases, which underlines the dependence on the orientation of the local polarization basis. To solve that problem, we define the coordinates properly by parallel transporting the reference axes from a chosen origin (the center of the patch, i.e. the center of a \texttt{HEALPix} superpixel at Nside = 4) to each pixel, thereby defining a common frame. This parallel transport provides the geometric foundation that makes the analysis of polarized dust foregrounds in terms of $Q$ and $U$ consistent and physically meaningful.

The validation patch is centered at $(l, b) = (213.7^{\circ}, -19.5^{\circ})$, where thermal dust emission clearly dominates over the nuisance. Henceforth, we will refer to this as the \emph{Orion patch}. The original Planck map was first rescaled to approximately match the SNR of the target patch to which the algorithm will ultimately be applied, and a nuisance realization from the corresponding ensemble was added to the rescaled image, thus creating a mock observation with a known ground truth. The initial rotated $Q$ maps, as well as the surrogate $d_Q$ map obtained after adding a nuisance realization, can be seen in Fig.~\ref{fig:Results_Q}.

\section{Formalism and algorithm} \label{sec:formalism}
\subsection{Principle of the algorithm}

For each polarization channel, the observed map for each Stokes parameter $a = Q,U$ is taken to be the sum of the respective thermal–dust emission \(s_a\) and a contamination term \(c_a\) that includes both the CMB residuals and the instrumental noise:
\begin{equation}
\label{eq:forward}
    d_a = s_a + c_a.
\end{equation}
Our goal is to construct maps \(\tilde{s}_a\) that are statistically consistent with the data once the effect of the contamination is taken into account. To do so, we will introduce an ensemble of constraints, estimated directly from the available data, that these maps have to fulfill. These constraints will be written using an ensemble of auto- and cross-statistics, which we will label $\Phi$, and whose choice is discussed in the following.

Following recent developments (\citealt{Regaldo2021}; \citealt{Delouis2022}; \citealt{Auclair2024}), we proceed by writing constraints in the statistics space that the prospective maps \(\tilde{s}_a\) must satisfy. The first of them are:
\begin{align} 
    &\Phi(d_a) \simeq \big\langle \Phi\big(\tilde{s}_a + c_a\big) \big\rangle_{c_a}, \label{eq:constraint_data}\\
    & \big\langle \Phi(c_a) \big\rangle_{c_a}  \simeq \Phi(d_a - \tilde{s}_a), \label{eq:constraint_nuisance}\\
    &\Phi(d_Q, d_U) \simeq \big\langle \Phi\big(\tilde{s}_Q + c_Q,\; \tilde{s}_U + c_U\big) \big\rangle_{c_Q,\,c_U}, \label{eq:constraint_cross}
\end{align}
where the average \(\langle \cdot \rangle_{c_a}\) is taken over the ensemble of contamination maps $\left\{c_{Q,i},c_{U,i}\right\}$ that is available, introduced in Sec.~\ref{sec:data}. Here, Eq.~\eqref{eq:constraint_data} imposes that $\tilde{s}_a + c_a$ is statistically compatible with the data $d_a$, on average over the ensemble of contamination samples. Similarly, Eq.~\eqref{eq:constraint_nuisance} imposes that $d_a - \tilde{c}_a$ is statistically compatible with the contamination model. Finally, Eq.~\eqref{eq:constraint_cross} requires that the $\tilde{s}_Q + c_Q$ and $\tilde{s}_U + c_U$ maps have cross-statistics which are statistically compatible with the data, in average over the ensemble of contamination samples. 

While these constraints require statistical independence between the $s_a$ maps and the contamination, an advantage is that they do not require any explicit knowledge of statistics involving only the $s_a$ maps, as $\Phi(s_a)$ or cross-statistics between $s_a$ and other maps. This framework thus allows us to efficiently use ancillary data, as long as they are independent, in the present case,

\clearpage
\begin{figure*}[p]
  \centering
  \includegraphics[width=\textwidth]{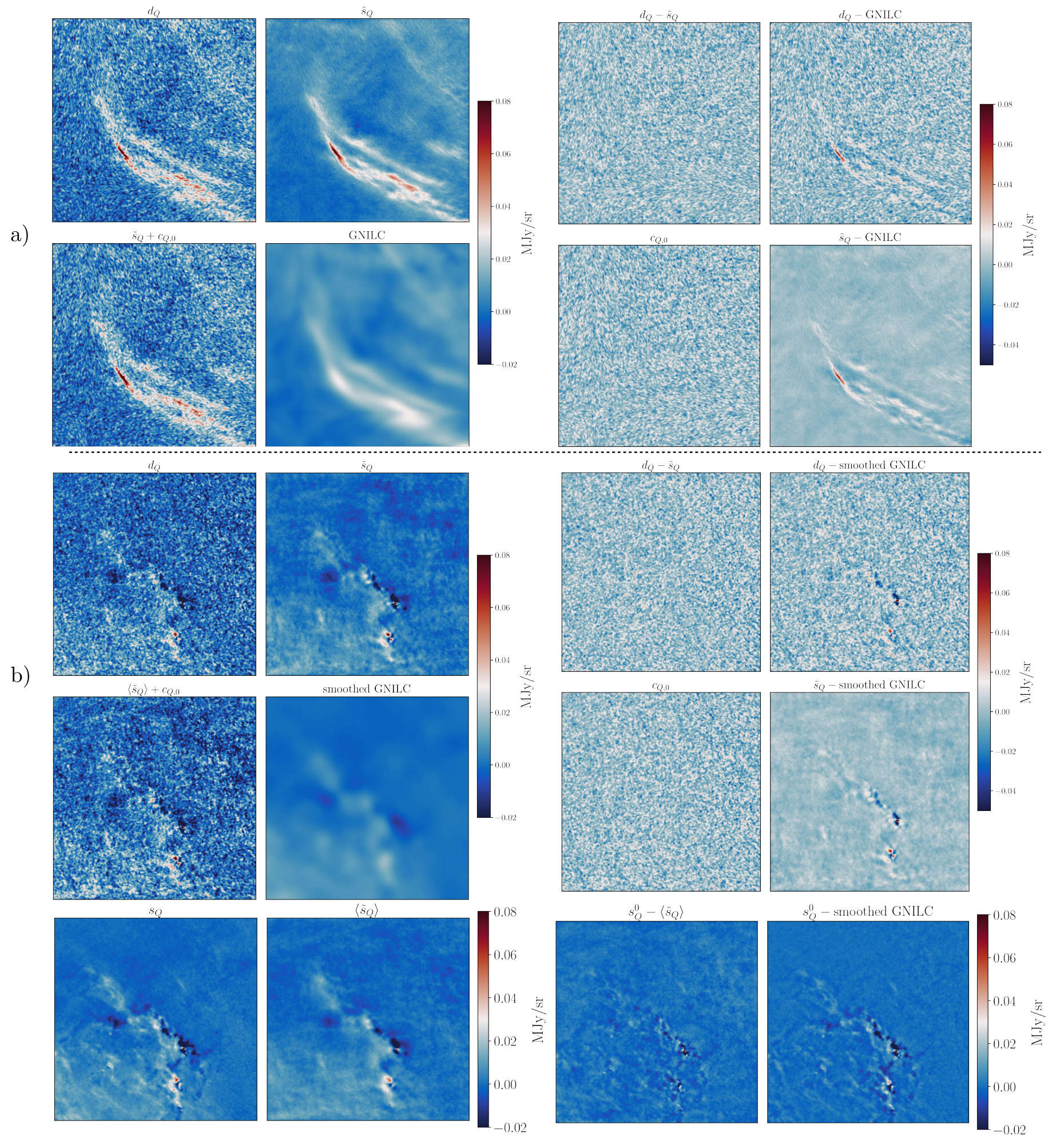}
    \caption{\textbf{Top:} Results of our component separation algorithm for the Stokes $Q$ component in the rotated polarization reference frame. The map labeled $d_Q$ is the CMB-subtracted Planck map, while $\tilde{s}_Q$ denotes the corresponding component-separated dust emission map. For comparison, the corresponding region of the GNILC map is shown. A random realization of the nuisance component, $c_{Q,i}$, is added to the $\tilde{s}_Q$ map to demonstrate statistical consistency with $d_Q$. Similarly, the recovered nuisance $d_Q - \tilde{s}_Q$ is compared with a random realization of the noise. In addition, we make plot $d_Q - \text{GNILC}$, in order to demonstrate the existence of dust residuals in the recovered nuisance map. Finally, the residual between $\tilde{s}_Q$ and the corresponding GNILC map is shown. \textbf{Bottom:} As above, but for the high-SNR region used for validation. Note that in this case the GNILC map is filtered to facilitate comparison with the results above. Finally, at the bottom tow we compare the true signal $s_Q$ with the average $\langle \tilde{s}_Q \rangle$ of ensembles obtained by using different initial conditions, as well as the residual between the true signal $s_Q$ and the recovered $\langle\tilde{s}_Q\rangle$, as well as with the recovered GNILC.}
  \label{fig:Results_Q}
\end{figure*}
\clearpage

\noindent to the contamination. In particular, we impose in this paper that \(\tilde{s}_a\) must reproduce the observed cross-statistics with the \(857\,\mathrm{GHz}\) intensity map \(d_I\):
\begin{equation} 
\label{eq:cross_con2}
    \Phi(d_a, d_I) \simeq \big\langle \Phi\big(\tilde{s}_a + c_a,\, d_I\big) \big\rangle_{c_a}.
\end{equation}
These constraints leverage the strong statistical dependency that is expected between these two signals, while not assuming a particular value for $\Phi(s_a,d_I)$. We note that such constraints could be used to leverage various types of ancillary data.

\vspace{-0.2cm}
\subsection{Choice of statistics and gradient descent optimization} \label{section3.2}

\paragraph{Gradient descent objective functions.}
The recovery of $\tilde{s}_a$ is carried out through a gradient–descent optimization in pixel space, whose objective function takes the form of a sum of seven loss functions, one for each of the constraints of Eqs.~\eqref{eq:constraint_data}–\eqref{eq:cross_con2} and each value of $a$. For example, the loss associated with the constraint given in Eq.~\ref{eq:constraint_data} for the Stokes $Q$ channel yields:
\begin{equation} 
\label{loss1}
    \mathcal{L}_1(u_Q) = \frac{1}{M} \sum_{i=1}^M|\Phi(u_Q + c_{Q,i}) - \Phi(d_Q)|^2.
\end{equation} 
These losses involve $u_a$ (\textit{i.e.}, $u_Q$ and $u_U$), the \emph{running maps} on which optimization is performed, the final values of which correspond to $\tilde{s}_a$ (\textit{i.e.}, $\tilde{s}_Q$ and $\tilde{s}_U$) after convergence of the gradient descent. The explicit form of the other individual loss functions is given in Eqs.~(\ref{eq:loss1_app}) - (\ref{eq:loss67}), in Appendix, and their purpose is schematically summarized in Table \ref{tab:losses}. The total objective function to be minimized is formed as the sum of all seven individual losses.

\begin{table}[h!]
\centering
\small 
\setlength{\tabcolsep}{5pt} 
\renewcommand{\arraystretch}{1.2} 
\begin{tabular}{c|ccccccc}
\hline
 Loss & 1 & 2 & 3 & 4 & 5 & 6 & 7 \\
\hline
Target & $d_Q$ & $d_U$ & $c_Q$ & $c_U$ & $d_Q \times d_U$ & $d_Q \times d_I$ & $d_U \times d_I$ \\
Variable & $u_Q$ & $u_U$ & $u_Q$ & $u_U$ & $u_Q \times u_U$ & $u_Q$ & $u_U$ \\
Eq. & (\ref{eq:constraint_data}) & (\ref{eq:constraint_data}) & (\ref{eq:constraint_nuisance}) & (\ref{eq:constraint_nuisance}) & (\ref{eq:constraint_cross}) & (\ref{eq:cross_con2}) & (\ref{eq:cross_con2}) \\
\hline
\end{tabular}
\vspace{0.20cm}
\caption{Summary of the seven losses used in the optimization. Each loss compares the statistics of a known target field (first row) to statistics estimated from specific running maps $u_a$ (second row) after either adding them to nuisance maps or subtracting them from $d_a$. The associated constraints are given in the third row.}
\label{tab:losses}
\end{table}

\vspace{-0.3cm}
In this paper, we perform this gradient descent starting from the data $d_a$, which lead to a single point-estimate $\tilde{s}_a$ for each Stokes parameter after convergence.
These maps, which verify constraints given Eqs.~\ref{eq:constraint_data} - \ref{eq:cross_con2}, are expected to accurately reproduce the statistical properties of $s_a$, at least at scales where the relative amplitude of the contamination is not too high. However, it should be noted that they do not necessarily reproduce the deterministic structures in $s_a$ at scales where the contamination is non-negligible (see~\cite{Regaldo2021,Delouis2022} for a discussion on this point).

We also produce a set of two $\langle \tilde{s}_a \rangle$ maps that are close to $s_a$ from a deterministic point of view (\textit{i.e.}, whose mean square error in pixel space is lower), which allows for a better comparison to GNILC. We do so by sampling additional maps close to the $\tilde{s}_a$ map, which verify constraints Eqs.~\ref{eq:constraint_data} - \ref{eq:cross_con2}, but whose structures differ at scales where the recovery is not deterministic anymore. These maps are obtained by the same algorithm as described above, but using $\tilde{s}_a + c_{a, i}$ as initial conditions, where $c_{a,i}$ are drawn from the nuisance ensemble. An ensemble average is then performed on these maps to produce $\langle \tilde{s}_a \rangle$, which is expected to be closer deterministically to the true map, as only the structures which are consistently reproduced along the different samples remain after averaging. It should be noted, however, that these maps do not verify anymore the constraints given Eqs.~\ref{eq:constraint_data} - \ref{eq:cross_con2}, since the structures at small scales are for instance smoothed.

While it seems clear from the results discussed below that the $\tilde{s}_a$ maps and $\langle \tilde{s}_a \rangle$ maps reproduce better the statistical and deterministic properties of $s$, respectively, a drawback of our approach is that we are currently unable to provide a consistent uncertainties estimates for these maps. This issue has been tackled in parallel for a similar framework, but with a mono-frequency and single-constraint problem~\citep{Pierre2026}. We let to future work to bring these studies together.

\vspace{-0.2cm}
\paragraph{Choice of statistics.}
Although, in principle, any choice of summary statistics $\Phi$ is possible, in practice they are selected to guaranty the stability of the pixel–space gradient descent, as well as to efficiently characterize the non-Gaussian features of the polarized foreground emission. As demonstrated in~\citet{Cheng2024} and~\citet{Mousset2024}, Scattering Covariance statistics have shown great promise in both regards. These form a family of ST statistics that are computed by calculating the covariances between the scattering coefficients computed at different oriented scales. For their mathematical definition and the explicit definition of $\Phi$, see Appendix~\ref{app:ScatCov}. 

Following these references, we adopt a normalization, estimated on the target side of the losses, that balances the relative contributions of large and small scales.
Indeed, complex multiscale physical processes typically exhibit a falling power-law behavior in the power spectrum, leading to high-frequency modes being subdominant in the loss if the coefficients are left unnormalized. To compensate for this, we normalize $\Phi$ so that all scales contribute with comparable weight. The specific normalization is shown in Appendix~\ref{app:ScatCov}. We note that for a single loss term, as was the case in~\citet{Regaldo2021}, the algorithm was found to perform better with a refined normalization putting more weight to the small scales. However, when multiple losses are combined simultaneously, it seems that the uniform normalization proposed in this paper gives very good results while being simple and stable.

\vspace{-0.2cm}
\subsection{Practical implementation}

In order to work with for $384\times384$ pixels maps, an ensemble of $J_{\max}=7$ absolute dyadic scales and $L=4$ orientations is used for all scattering–covariance computations (see Appendix~\ref{app:ScatCov} for definitions). This choice fixes the number of coefficients entering each loss term: for single–field statistics $\Phi(x)$, the total number of coefficients is $5\,011$, while for cross–field statistics $\Phi(x_1,x_2)$, it is $17\,960$. These values determine the dimensionality of the constraints appearing in each of the seven losses and thus in the objective function.

The optimization is performed in \texttt{PyTorch} using a limited-memory Broyden–Fletcher–Goldfarb–Shanno (L--BFGS) optimizer \citep{Nocedal1980}. At each iteration, every loss term is evaluated over a random batch of $10$ nuisance realizations, yielding a stochastic approximation of the expectation over contamination while maintaining the efficiency of a quasi--Newton update. Optimization is stopped after $35$ steps for each iteration. For $384\times384$ maps, the gradient--descent procedure converges in approximately $20$ minutes on an NVIDIA A100 80 GB GPU. In order to compute the ensemble maps $\bar s_a$, we use $33$ samples for both the Northern patch and Orion patch.

\section{Results} \label{sec:results}

\vspace{-0.1cm}
\subsection{North patch}

\vspace{-0.1cm}
In Fig. \ref{fig:Results_Q} we present the results of our component-separation method applied to the North patch. We clearly see in it that the recovered map $\tilde{s}_Q$ reconstructs small-scale non-Gaussian structures beyond the noise level. 
In contrast, the GNILC map of this region appears to smooth out fine-scale fluctuations, illustrating the stronger filtering applied by that method. 
By visual inspection, one also observes that the composite maps 
\(\tilde{s}_Q + c_{Q,i}\) closely resemble \(d_Q\), and that the residuals 
\(d_Q - \tilde{s}_Q\) similarly resemble the nuisance realizations \(c_Q\), 
indicating good statistical proximity in both comparisons. 
For each of these comparisons, we emphasize that while visual agreement is not a sufficient condition, achieving it is a very good indicator of similar non-Gaussian structures.

The power spectra corresponding to the previous comparisons are shown in Fig. \ref{fig:PS}. These spectra were computed on apodized maps and binned to reduce statistical variance at high-$k$ modes. 
At the spectral level, we observe that the power spectrum of $\langle \tilde{s}_Q + c_Q \rangle$ agrees with that of $d_Q$ within the one sigma level estimated from the nuisance variability. Further, the power spectrum of $d_Q - \tilde{s}_Q$ also agrees with that of the nuisance maps within one sigma of the nuisance. 
The power spectrum of \( \tilde{s}_Q \) is also compared with that of the polarized dust emission inferred from the cross-correlation of two half-ring observations of the same sky patch. Within the apparent variance of the half-ring cross-spectrum, the recovered \( Q \)-component power spectrum shows good agreement. Finally, the power spectrum of the corresponding GNILC map is shown for reference. 

In Fig.~\ref{fig:PySM}, we compare the reconstructed dust polarization map obtained from our method with three state-of-the-art simulated dust models from PySM 3 (\cite{PySM}; \cite{PySM3}; \cite{Borrill_2025}) corresponding to increasing levels of physical complexity. Models d9, d10, and d12 represent, respectively, a single–modified blackbody model with fixed parameters, a spatially varying modified blackbody model, and a six independent layers dust model that accounts for multiple dust populations along the line of sight \citep{2018MNRAS.476.1310M}. From this comparison, it appears that our method is able to recover non-Gaussian structures that are more compatible with the data than the previous models.

\vspace{-0.4cm}
\subsection{Validation on the Orion patch}

To validate our method, we applied the algorithm to the rescaled Orion patch (see section \ref{sec:patches}), 
Figure~\ref{fig:Results_Q} shows the resulting output maps, similarly than for the North patch. The map $\tilde{s}_Q + c_{Q,i}$, seems to agree with the mock observation $d_Q$, suggesting statistical consistency between the reconstructed and observed map. 
Similarly, the residual map $d_Q - \tilde{s}_Q$ agrees well with the true nuisance $c^0_Q$, confirming that the recovered nuisance component captures the expected statistical properties. 

In addition, a visual comparison between the recovered map $\tilde{s}_Q$ and the ground truth $s_Q^0$ indicates that the contamination has been effectively removed while preserving the non-Gaussian structure of the polarized dust emission, even well below the noise level, while not recovering certain small scale features below a given angular scale (i.e., at higher $\ell$), especially those not close of the large central structure. Note that in this map, the GNILC map has been filtered so that it contains power in the same scales as in the GNILC map for the North patch. 

Finally, the bottom row of Fig.~\ref{fig:Results_Q} shows a comparison between the true $s_q$, $\langle \tilde{s}_Q \rangle$, and GNILC. Overall, $\langle \tilde{s}_Q \rangle$ provides a better small-scale deterministic reconstruction, as seen in the maps and their differences from the truth.

\begin{figure}[t!]
  \centering
  \includegraphics[width=0.45\textwidth]{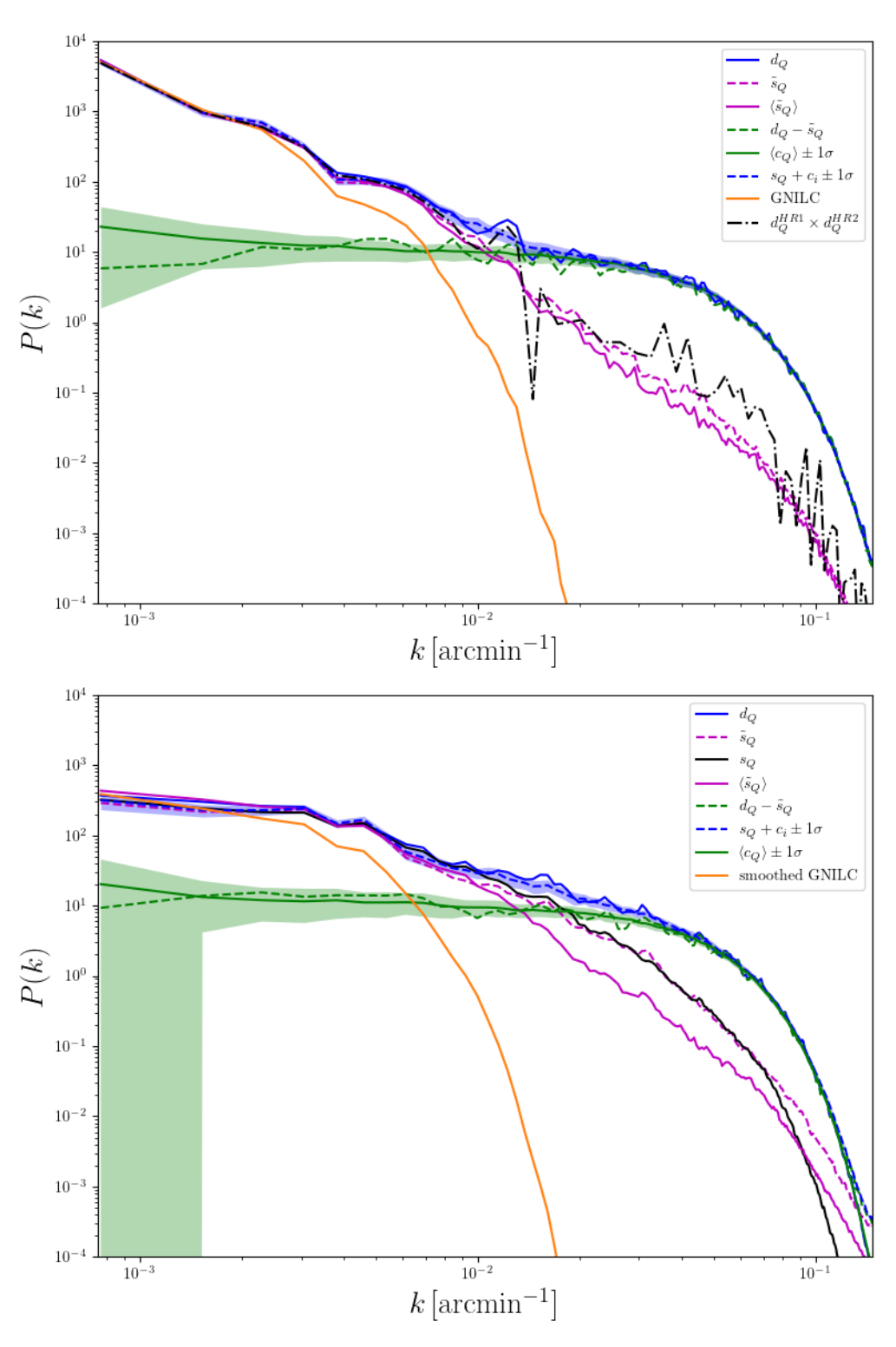}
  \caption{\textbf{Top:} Power spectra of the maps shown in Fig. \ref{fig:Results_Q}a. 
The solid blue line shows the initial Planck map $d_Q$, while the dashed blue line (with $1\sigma$ margin) corresponds to the recovered map after adding the nuisance. 
The dashed purple line correspond to the recovered dust signal $\tilde{s}_Q$. For comparison, the power spectrum of the true polarized dust emission is approximated by taking the cross-spectrum of the two half-ring maps (dashed black-dotted line). The solid purple line corresponds to the average recovered signal $\langle \tilde{s}_Q \rangle$ (see Sec.~\ref{section3.2}).
Finally, the solid and dashed green lines correspond to the nuisance (with $1\sigma$ margin) and recovered nuisance maps. 
For comparison, the power spectrum of the filtered GNILC map is also shown in orange. \textbf{Bottom:} As above, but for the maps of Fig. \ref{fig:Results_Q}b. In this case, the power spectrum of the true polarized dust emission (solid black line) is known.
\vspace{-0.6cm}}
  \label{fig:PS}
\end{figure}

For this validation patch as for the North patch, Fig.\ref{fig:PS} shows the power spectra of the different maps. 
We find that the power spectra of the recovered polarized dust emission, $\tilde{s}_Q$, and that of the true map $s_Q$ are consistent across scales, 
demonstrating the success of our component-separation even at scales where the contamination dominates by between one and two orders of magnitude. 
As in the application to the North Patch, the power spectrum of $\tilde{s}^0_Q + c_Q$ agrees with the data at all scales within one sigma due to the nuisance variability, as is the case with the spectra of $\tilde{c} = d - \tilde{s}$ and $c$. 
Fig.~\ref{fig:cross_spectra} in Appendix presents various cross-spectra comparing our recovered maps — after the addition of nuisance realizations — with those derived directly from the Planck data, illustrating the statistical consistency of the recovered maps across scales, at least at the two-point level. 

Note that, in this validation, the true contamination is drawn from the same model used for component separation. When working with real data, as with the North patch, model misspecification may also occur.


\vspace{-0.2cm}
\begin{figure}[t!]
    \centering
    \includegraphics[width=1\linewidth]{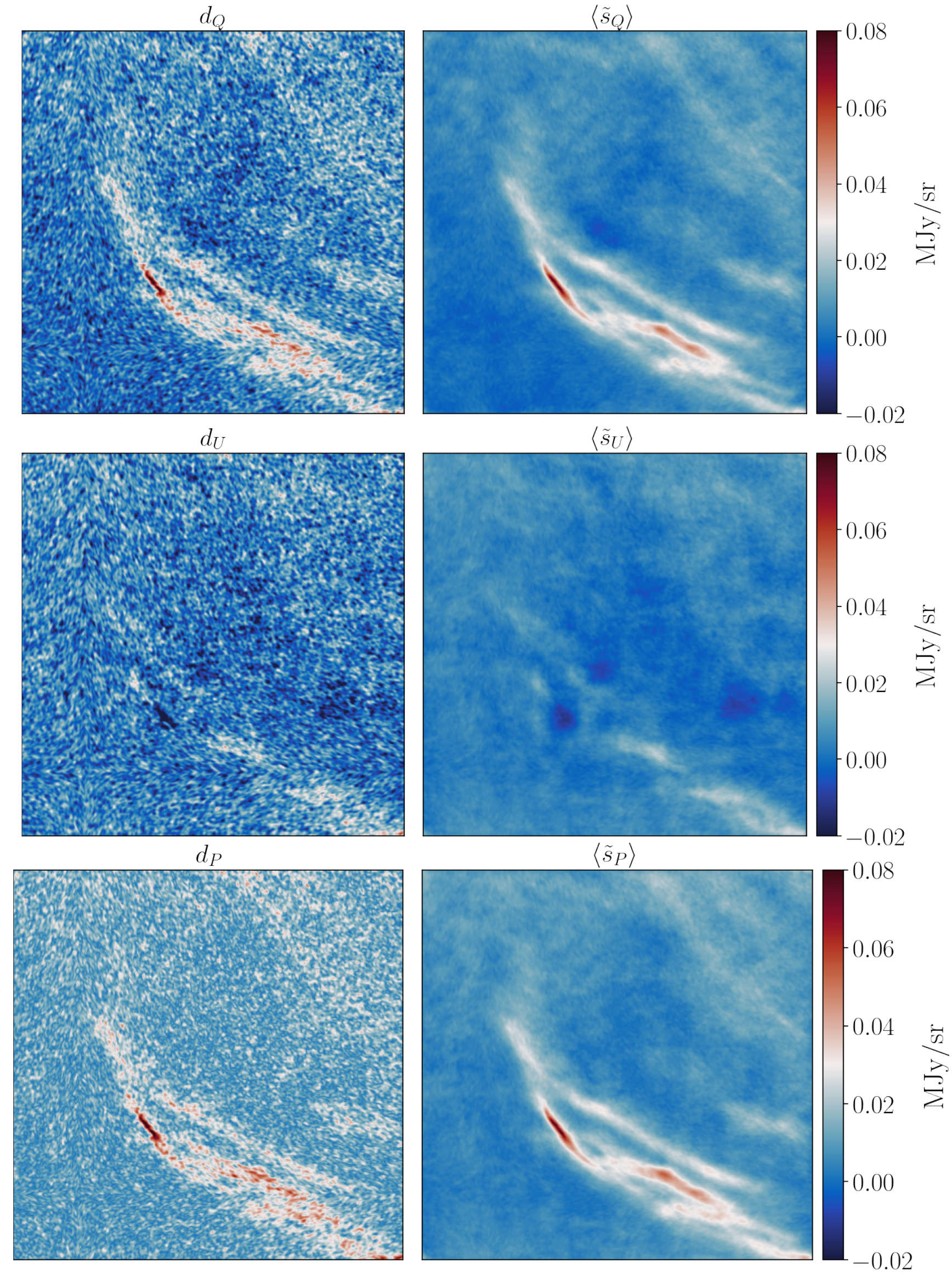}
    \caption{Results for the North patch. \textbf{Top}: observed $d_Q$ component (left), and reconstructed $\langle s_Q \rangle$ (right). \textbf{Middle and Bottom}: As above but for the $U$ and polarized intensity $P$ components, respectively. 
    \vspace{-0.5cm}}
    \label{fig:north_patch_results}
\end{figure}


\section{Conclusions} \label{sec:conc}

In this work, we have presented a multi-constrained component separation framework based on Scattering Transform (ST) statistics for the recovery of polarized Galactic foregrounds from Planck data. The method reconstructs maps of polarized dust emission whose ST statistics are consistent with those expected from the Planck data, once a realization of the nuisance (noise and CMB) is added to it. These maps can be used to study the statistical properties of the dust polarzied emission, or even to directly construct a ST model of this signal.

Two different sets of $Q$ and $U$ maps are produced, two $\tilde{s}_a$ maps that reproduce well the expected statistical properties, and two $\langle \tilde{s}_a \rangle$ maps that are a better estimate of deterministic structures, but which are slightly filtered at scales where the contamination is non-negligible. Beyond meeting the constraints of visual statistical consistency, our results show that the $\tilde{s}_a$ maps remains consistent across further power and cross spectra diagnostics, retaining small-scale information that is lost in the GNILC map. In the validation test, our recovered map seems to reproduce the statistical properties of the true map even at angular scales where the dust amplitude lies several orders of magnitude below the noise. We also observe that on a pixel-based comparison, the $\langle \tilde{s}_a \rangle$ maps seem to consistently improve on GNILC. These results demonstrate the significant potential of our approach compared to the current state of the art when applied to real data. While we are currently not able to provide uncertainties for the results produced, we note that this task has been undertaken in a parallel project~\citep{Pierre2026}.

A major strength of this framework is that it is inherently flexible and can be generalized in several directions. In future work, one could extend the method to perform multi-frequency component separation in polarization, thereby capturing the cross-frequency statistical dependency of Galactic dust emission. With accurate uncertainty estimates, this could enable the production of a distributions of high quality multi-frequency polarized microwave dust models.
Beyond Planck, the approach could also be extended to other current and upcoming CMB datasets, such as ACT and SO. More broadly, the generality of the scattering-based formalism makes it applicable to a wide class of inverse and generative problems involving complex non-Gaussian physical fields.

\begin{acknowledgements}
We sincerely thank  S. Clark, J.-M. Delouis, F. Levrier,  S. Ghosh, I. Grenier, S. Mallat, and S. Pierre, for various insights and advice. 
The authors acknowledge Interstellar Institute's program "II7" and the Paris-Saclay University's Institut Pascal for hosting fruitful discussions behind this work.
This work received government funding managed by the French National Research Agency under France 2030, reference numbers “ANR-23-IACL-0008” and “ANR-25-CE46-6634".
This work was granted access to the HPC resources of MesoPSL financed by the Region Ile de France and the project Equip@Meso (reference ANR-10-EQPX-29-01) of the programme Investissements d’Avenir supervised by the Agence Nationale pour la Recherche.
This research used resources of the National Energy Research Scientific Computing Center (NERSC), a Department of Energy User Facility (project mp107d-2025).
\end{acknowledgements}

\vspace{-0.8cm}

\bibliographystyle{aa}   
\bibliography{bibliography}

\appendix

\section{HEALPix pixel square patches} \label{app:HEALPix}
We extract from full sky HEALPix maps square patches centered on the centers of HEALPix pixels at resolution Nside = 4, directly as the set of healpix subpixels of each given superpixel. Although the output data are in the format of squares, there is no reprojection from the original HEALPix pixelization. The drawback is that the output square maps are ``distorted", as HEALPix pixels at Nside = 4 are not strictly-speaking ``square'' (even if they are in the format of a square grid). The main advantage of making ``pseudo squares'' that match HEALPix pixels is that we avoid any reprojection effects.

The objective of extracting such patches is to use data processing tools designed for square images. In particular, we will make use of Fourier transforms, wavelet transforms, convolutions, and filtering on square maps. Eventually, these maps could be recombined later on into a full spherical map.

In order to avoid discontinuities, border effects, and aliasing, we add bordering pixels to each Nside = 4 superpixel (of size 512$\times$512 for original maps at Nside = 2048). Here, we adopt 128-pixel wide borders. The dimensions of the final extracted maps is hence 768$\times$768 pixels. The additional padding with 128-pixel wide borders allows for apodization outside of the area of interest for the patch, and taking care of border effects if when we recombine the patches into a full sky. 

In contrast to the CMB, where the $E$- and $B$-mode decomposition is the natural language for cosmological inference, Galactic foregrounds such as thermal dust emission are best studied in terms of the Stokes parameters $Q$ and $U$. Indeed, $Q$ and $U$ are the directly measured, local quantities that retain a simple physical interpretation: they encode the polarization amplitude and angle relative to a tangent-plane basis at each point on the sky, which are directly connected to the geometry of the Galactic magnetic field and dust grain alignment. In comparison, the $E/B$ decomposition is intrinsically non-local, involving harmonic transforms that mix information across large regions of the sky. While this is optimal for characterizing CMB fluctuations, it obscures the local structure and non-Gaussian statistics that are crucial for understanding the physics of dust polarization. Note that for instance, the $E$ and $B$ fields for a single bright pixel vanish in that pixel, making $IE$ and $IB$ correlations equal to zero, while $IQ$, $IU$, and $QU$ correlations are non-zero for a non-vanishing polarization fraction when none of $I$, $Q$ and $U$ vanishes.

There is, however, a practical difficulty. Working in $Q$ and $U$ introduces the geometric subtlety that their definition depends on the orientation of the local polarization basis, which varies across the sphere. To compare polarization measurements across pixels, it is therefore necessary to enforce a consistent choice of basis locally in the patch of interest. This is accomplished by parallel transporting the reference axes from a chosen origin (e.g.\ the center of a \texttt{HEALPix} superpixel at Nside = 4) to each pixel, thereby defining a common frame. The relative rotation angle $\psi$ between the \texttt{HEALPix} convention at each pixel and the transported frame specifies how $Q$ and $U$ transform. Correcting for this angle reduces spurious mixing of the Stokes parameters and ensures that spatial correlations in $Q$ and $U$ reflect genuine astrophysical patterns rather than coordinate artifacts. In this way, parallel transport provides the geometric foundation that makes the analysis of polarized dust foregrounds in terms of $Q$ and $U$ both consistent and physically meaningful.

To consistently define polarization across a patch, we parallel transport a reference axis from the patch center to each pixel. Let $\hat{\boldsymbol v}$ be the transported reference axis and $\hat{\boldsymbol e}_\theta, \hat{\boldsymbol e}_\Phi$ the local tangent-plane basis at a pixel. The rotation angle $\psi$ between the transported axis and the local basis is defined by projecting $\hat{\boldsymbol v}$ onto the local axes:
\[
\psi = \operatorname{atan}\big(\, \hat{\boldsymbol e}_\Phi \cdot \hat{\boldsymbol v} \,,\, \hat{\boldsymbol e}_\theta \cdot \hat{\boldsymbol v} \,\big).
\]
This angle $\psi$ is then used to rotate the Stokes parameters $(Q,U)$ from the local HEALPix frame into the common, transported frame via the spin-2 transformation:
\[
Q' = Q\cos(2\psi) + U\sin(2\psi), \qquad
U' = -Q\sin(2\psi) + U\cos(2\psi),
\]
where $Q'$ and $U'$ represent the rotated Stokes parameters.
After this rotation, all pixels in the patch share a consistent polarization frame, allowing the most meaningful comparison and analysis of $Q$ and $U$ across the patch. 



\vspace{-0.3cm}
\section{Input data and nuisance} \label{app:input_data}

We have chosen to work with the PR4 Planck NPIPE maps~\cite{npipe_2020}, as they have higher signal-to-noise at intermediate scales in polarization. We process the 353~GHz frequency $Q$ and $U$ maps in the following way: we subtract a Wiener-filtered CMB map and convolve the maps to a common resolution of 10~arcmin. We assume that the input polarization maps have a Gaussian beam with a FWHM of 4.76~arcmin~\cite{npipe_2020}. We convert the frequency maps to MJy/sr. 

The nuisance maps for 353~GHz $Q$ and $U$ maps are the sum of noise and CMB residuals. We use the difference of simulated half-ring maps as the noise proxy, as they share the same systematics and have uncorrelated noise. We borrow 100 realizations of half-ring maps from the NPIPE simulations at NERSC\footnote{The simulations can be found at NERSC: /global/cfs/cdirs/cmb/data/planck2020}. CMB residual is estimated by propagating the Wiener filter coefficient to simulated CMB maps as:
\begin{equation}
    a_{{\ell m}_{{\rm res}_i}} = a_{{\ell m}_{{\rm CMB}_i}} - w a_{{\ell m}_{{\rm CMB}_i}}
\end{equation}
where $i$ represents mode \{T, E, B\}, $a_{{\ell m}_{{\rm  res}_i}}$ is the residual CMB harmonic coefficient, $a_{{\ell m}_{{\rm CMB}_i}}$ is the CMB harmonic coefficient, and $w$ is the Wiener filter coefficient.
As the 857~GHz intensity map is only used as a tracer of dust emission in cross-statistics with 353~GHz Q and U maps, and considering that its contamination by CIB and noise is expected to be independent of both signal and noise at 353~GHz, we simulate only nuisance maps for the 353~GHz Q and U maps.

\begin{figure*}[t!]
  \centering
  \includegraphics[width=\textwidth]{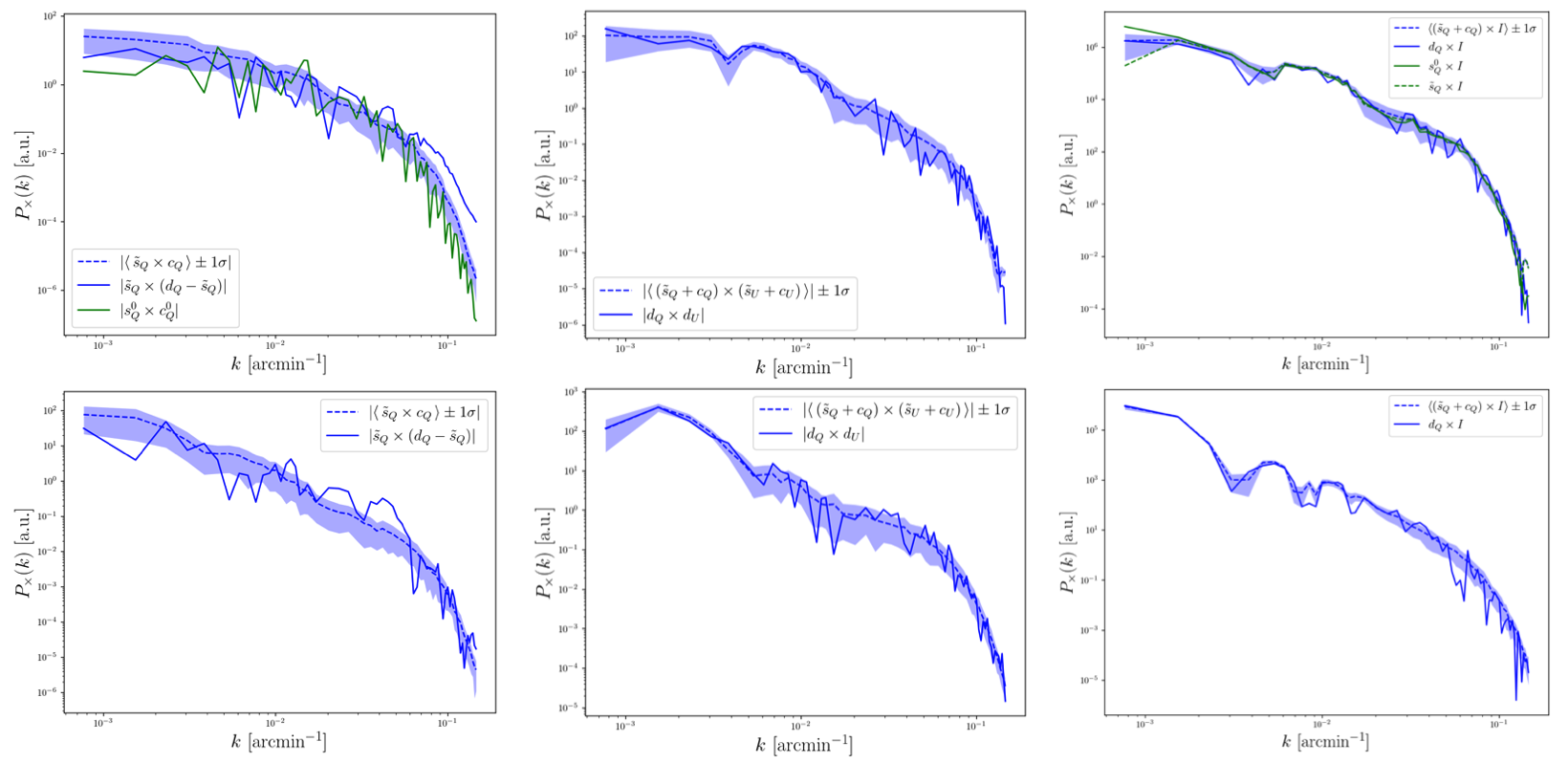}
  \caption{Various cross spectra between the recovered maps and the expected spectra from the data. \textbf{Top}: Validation on the Orion region (true map indicated by $s_Q$). \textbf{Bottom:} application to the North patch.}
  \label{fig:cross_spectra}
\end{figure*}

\vspace{-0.3cm}
\section{Scattering covariances} \label{app:ScatCov}

The set of summary statistics that we will use are scattering covariances. Suppose that $j$ and $\theta$ define a specific dyadic scale and an orientation of a given map $x$. Then, $\lambda \equiv (2^{-j}, \theta)$ defines a specific oriented scale. If $\psi^\lambda$ is a band-pass filter at oriented scale $\lambda$, then scattering covariances are defined as:

\begin{equation} \label{eq:stats}
    \begin{aligned}
S_1(\lambda_1) &\equiv \langle | x \star \psi^{\lambda_1} | \rangle \\
S_2(\lambda_1) &\equiv \langle | x \star \psi^{\lambda_1} |^2 \rangle \\
S_3(\lambda_1, \lambda_2) &\equiv \frac{\text{Cov} \big[ x \star \psi^{\lambda_2}, \, | x \star \psi^{\lambda_1} | \star \psi^{\lambda_2} \big]}{\sqrt{\langle | x \star \psi^{\lambda_1} |^2 \rangle} \sqrt{\langle | x \star \psi^{\lambda_2} |^2 \rangle}} \\
S_4(\lambda_1, \lambda_2, \lambda_3) &\equiv \frac{\text{Cov} \big[ | x \star \psi^{\lambda_1} | \star \psi^{\lambda_3}, \, | x \star \psi^{\lambda_2} | \star \psi^{\lambda_3} \big]}{{\sqrt{\langle | x \star \psi^{\lambda_1} |^2 \rangle} \sqrt{\langle | x \star \psi^{\lambda_2} |^2 \rangle}}},
\end{aligned}
\end{equation}
where $\star$ denotes a convolution, and the averages are taken over the pixel values. Therefore we can define the quantity

\begin{equation} 
    \Phi(x) \equiv \{\mu/\sigma, \log(S_1), \, \log(S_2), \, S_3, \, S_4 \}.
\end{equation}
where $\mu$ and $\sigma$ denote the pixel-wise mean and standard deviation of $x$. Notice that the terms $S_2$, $S_3$, and $S_4$ can take two different maps $x_1$ and $x_2$ as input, for instance $S_2(\lambda_1) = \langle ( x_1 \star \psi^{\lambda_1} ) ( x_2 \star \psi^{\lambda_1} )^* \rangle$ and vice versa. Therefore, we can in general write the cross-statitics maps $\Phi(x_1, x_2)$ for $x_1 \neq x_2$ and define $\Phi(x, x) \equiv \Phi(x)$.

The normalization in equation is justified as follows; without the normalization, high frequency modes will be completely subdominant compared to low-frequency ones in the objective function, and so the gradient descent will be only driven by large scales. With the normalization included, a reweighting is performed such that small scales modes are not neglected. 

Notice, finally, that since the logarithm is taken for the $S_1$ and $S_2$, as this definition was found to aid convergence, no normalization is required for these two terms. 

The seven loss functions minimized in this work to recover the maps 
$\{\tilde{s}_Q, \tilde{s}_U\}$, which satisfy the constraints given in 
Eqs.~\ref{eq:constraint_data}--\ref{eq:cross_con2}, are defined as follows:
\begin{equation} \label{eq:loss1_app}
    \mathcal{L}_{1,2}(u_a) = \frac{1}{M} \sum_{i=1}^{M}
    \big|\Phi(u_a + c_{a,i}) - \Phi(d_a)\big|^2,
\end{equation}
\vspace{-0.6cm}
\begin{equation}
    \mathcal{L}_{3,4}(u_a) =
    \big|\Phi(d_a - u_a) - \frac{1}{M}\sum_{i=1}^{M} \Phi(c_{a,i})\big|^2,
\end{equation}
\vspace{-0.6cm}
\begin{equation}
    \mathcal{L}_5(u_Q, u_U) = \frac{1}{M} \sum_{i=1}^{M}
    \big|\Phi(u_Q + c_{Q,i},\, u_U + c_{U,i}) - \Phi(d_Q, d_U)\big|^2,
\end{equation}
\vspace{-0.6cm}
\begin{equation} \label{eq:loss67}
    \mathcal{L}_{6,7}(u_a, d_I) = \frac{1}{M} \sum_{i=1}^{M}
    \big|\Phi(u_a + c_{a,i},\, d_I) - \Phi(d_a, d_I)\big|^2.
\end{equation}
Here, $a$ indexes the polarization channels ($a \in \{Q, U\}$), 
and $M$ denotes the number of noise realizations used to estimate the expected 
scattering statistics.

\section{Additional figures}

\begin{figure*}[b!]
  \centering
  \includegraphics[width=1\textwidth]{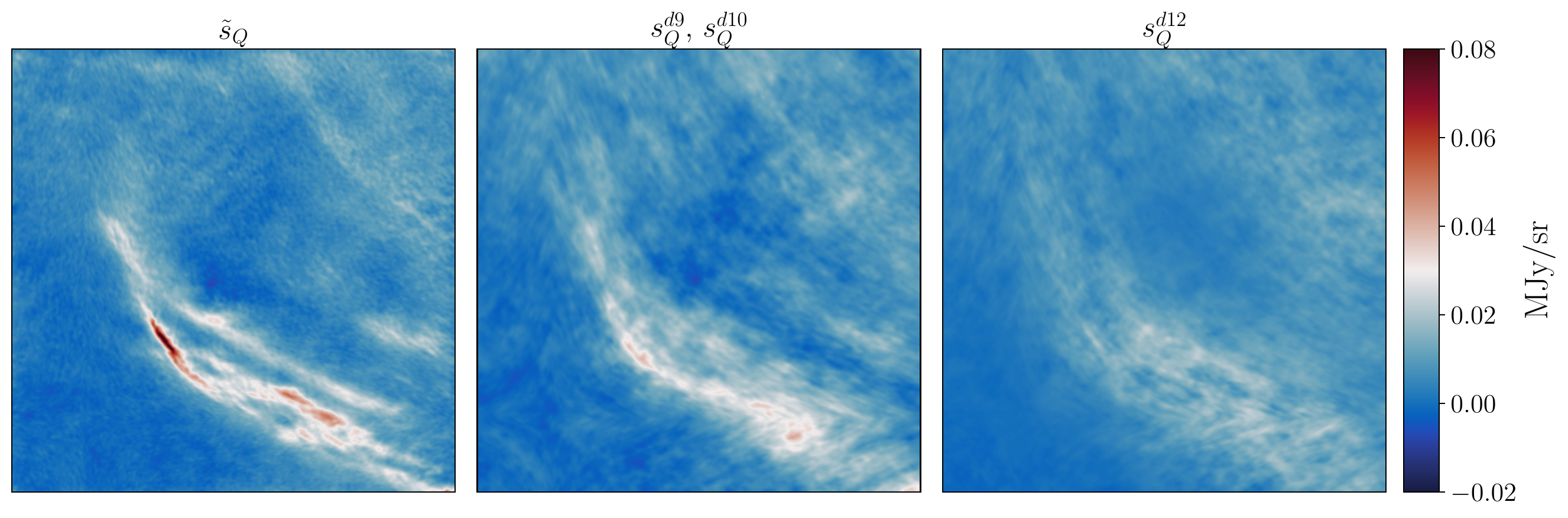}
  \caption{Comparison of our recovered model with the corresponding PySM 3 maps for the same region. The maps $\tilde{s}^{d9}_Q$, $\tilde{s}^{d10}_Q$, and $\tilde{s}^{d12}_Q$ correspond to three different models provided by PySM 3: a single–modified blackbody model with fixed spectral parameters (d9), a spatially varying modified blackbody model (d10), and a six independent layers dust model, accounting for multiple dust populations (d12), respectively.}
  \label{fig:PySM} 
\end{figure*}

Fig.~\ref{fig:cross_spectra} show the cross-spectra of the different maps for both the North patch and the Orion region. Fig.~\ref{fig:PySM} compare our $\tilde{s}_Q$ map for the North Patch with different PySM 3 models.

\end{document}